\documentclass[aps,pre,twocolumn,superscriptaddress]{revtex4}
 \usepackage{amsmath,graphicx,color,epsfig}
 \DeclareMathOperator{\mode}{mode} 
 
\begin{document}
 \title{Coevolutionary dynamics with global fields}
 \author{M. G. Cosenza}
 \affiliation{School of Physical Sciences \& Nanotechnology, Universidad Yachay Tech, Urcuqu\'i, Ecuador}
 \affiliation{Universidad de Los Andes, M\'erida, Venezuela}
 \author{J. L. Herrera-Diestra}
 \affiliation{Department of Integrative Biology, University of Texas at Austin, TX 78712, U.S.A.}
\affiliation{Centro de Simulaci\'on y Modelos (CeSiMo), Universidad de Los Andes, M\'erida, Venezuela}

 \begin{abstract}
	We investigate the effects of external and autonomous global interaction fields on an adaptive network of social agents with an opinion formation dynamics. 
	The model represents a society
	subject to global mass media such as a directed opinion influence or a feedback of endogenous cultural trends. 
	We show that, in both situations,  mass media always contribute to consensus and to prevent the fragmentation of the social network. We present a discussion of these results.
\end{abstract}
 \date{Entropy  \textbf{24}(9), 1239 (2022)}
 \maketitle


	\section{Introduction}
	Many physical, chemical, biological, social, and economic systems are subject to global interactions. A global interaction in a system occurs when all its constituents share a common influence or source of information \cite{Kaneko}. The origin of a global interaction can be either external, as in a forcing field; or autonomous, such as a mean field or a feedback coupling function that depends on the elements of the system \cite{Parra,Piko}. Global interactions appear, for example, in parallel electric circuits, coupled oscillators \cite{Kuramoto,Naka}, Josephson junction {arrays}~
	\cite{Wie}, charge density waves \cite{Gruner}, multimode lasers \cite{Wie2}, neural networks, evolution models, ecological systems \cite{Kan1}, social networks \cite{Newman}, economic exchange \cite{Yako}, mass media influence \cite{Media1,NJP,Cross}, and cultural globalization \cite{Media2}. A complete graph or fully connected network, where any node can interact each each other, can be seen as a global interaction.
	Diverse collective behaviors can emerge in globally coupled oscillators, such as complete and generalized chaos synchronization, dynamical clustering, nontrivial collective behavior, chaotic itinerancy, quorum sensing, and chimera states  \cite{Kan2,Manrubia,Sethia,Yedel,Ojalvo,Wang,Monte,Taylor,Tinsley,Scholl}. 
	Systems possessing coexisting global and local interactions have also been studied \cite{Kan3}.

	Most of the research on the effects of global interaction fields has been conducted considering the evolution of the states of the nodes on a fixed network. However, many complex systems observed in nature can be described as dynamical networks of interacting elements where the states of the elements and their connections influence each other and evolve simultaneously \cite{Eguiluz1,Rohl,Eguiluz2,Blasius,Sayama}. The terms coevolutionary dynamical system and adaptive network~\cite{Eguiluz2,Blasius} have been employed for systems exhibiting this coupling between the network topology and node state dynamics. Coevolution models have been studied in spatiotemporal dynamical systems, such as neural networks \cite{Ito,Gross}, coupled map lattices~\cite{Ito2,Gong}, motile elements \cite{Shibata}, synchronization in networks \cite{Arenas}, as well as in spin dynamics \cite{Mandra}, epidemic propagation \cite{Gross2,Risau,Shaw}, game theory \cite{Eguiluz1,Eguiluz2,Gao}, and also in the context of social dynamics, such as opinion formation and cultural polarization \cite{holme,Min1,20,22,23,24,Herrera,Gargi}. Coevolutionary systems usually exhibit a transition between two network configurations: a large connected graph where most nodes share the same state, and a fragmented network of small disconnected components, each composed by nodes in a common state \cite{holme,Min1,20,22,23,24,Herrera,Gargi}. This network fragmentation transition is related to the difference in time scales of the processes that characterize the two dynamics: the state of the nodes and the network of interactions~\cite{holme}. 
	
	In this article we investigate the effects of global interaction fields on coevolutionary dynamics. 
	We study the competition between adaptive rewiring and global interactions in a network.
	We consider external global fields or autonomous global fields acting on an adaptive network of social agents with an opinion formation dynamics based on a simple imitation rule. In the context of social phenomena, our system can be considered as a model for a society subject to global mass media that represent a directed opinion influence or a feedback of endogenous cultural trends. We show that, in both situations, global mass media contribute to consensus and to inhibit the fragmentation of the social network. We present a discussion of these results.
	
	\section{Methods}
	
	We consider a population of $N$ social agents represented as nodes on an initially random network of the Erdos--Renyi type with an average degree $\langle  k \rangle$, i.e., $\langle  k \rangle$ is the average number of edges per node \cite{Network}. We denote by $\nu_i$ the set of $k_i$ neighbors of node $i$ (\mbox{$i=1,2,\ldots,N$}). We let $g_i$  be the state variable or the opinion of agent $i$, where $g_i$ can take any of the $G$ equivalent options in the set $\{1,2,\ldots,G\}$; i.e., we assume that the states of the nodes were discrete. 
	
	We introduce a global field $\Phi$ that can interact with all the elements in the system and that has a state $g_\Phi\in\{1,2,\ldots,G\}$. The global field can be interpreted as an additional neighbor shared by each node $i$ with whom an interaction is possible. Then, the network subject to the global field $\Phi$ corresponds to a dynamical system possessing both local and global interactions. 
	
	We consider two types of global fields $\Phi$:
	
	(i) An \textit{external global field}
    whose value $g_\Phi$ is chosen from the set $\{1,\ldots,G\}$ and remains fixed during the evolution of the system. The external field corresponds to a constant spatially uniform influence acting on the system. A constant external field can be interpreted as a specific state (such as an opinion, a message, or advertisement) being transmitted by
	mass media over all the elements of a social system.
	
	(ii) An \textit{autonomous global field} whose value $g_\Phi$ depends on the state variables of the elements in the system. 
	Here, we define $g_\Phi$ as the statistical mode of the distribution of state variables of the agents in the system at a given time, denoted as \mbox{$g_\Phi=\mode \{ g_1, g_2, g_3,\ldots, g_N\}$}. That is, we assign $g_\Phi$ as the most abundant value exhibited by states of all the nodes in the system at a given time. If the maximally abundant value was not unique, one of the possibilities is chosen at random with equal probability. The autonomous field is spatially uniform, but its value may change as the system evolves. In the context of opinion or cultural models, this field may represent an endogenous global mass media influence that transmits the predominant opinion, cultural trend, or fashionable behavior present in a society.
	
	We characterize the intensity of a global field by a parameter $B \in [0,1]$ that expresses the probability of interaction of any agent with the field. Then, the probability of interaction between two agents is proportional to $(1-B)$. On the other hand, the rewiring process in the network takes place with a probability $P_r$, and we assume that the node dynamics occurs with probability $1-P_r$. Thus, the node dynamics is coupled to the rewiring dynamics giving rise to a coevolutionary system. For the node state dynamics, we
	implemented a voter-like model that has been employed in 
	coevolution of opinions and networks \cite{holme} and in various situations \cite{Min1,20,22,23,24,Herrera}. 
	
	We build the initial random network with parameter values $N=1000$ and \mbox{$\langle  k \rangle=4$}. Then, the states $g_i$ are assigned to the nodes at random with a uniform distribution. Therefore there are, on average, $N/G$ agents in each state in the initial network. Here, we fixed the number of options at the value $G=100$.
	
	Then, the coevolution dynamics of the system subject to a global field $\Phi$, either external or autonomous, is defined by the following iterative algorithm:
	\begin{enumerate}
		\item Choose at random an agent $i$ such that $k_i>0$.
		\item With probability $P_r$, select at random an agent
		$j \in \nu_i$ and another agent $l \notin \nu_i$ such that $g_i=g_l$;
		remove the edge $(i,j)$ and set the edge $(i,l)$.
		\item With probability $(1-P_r)B$, set $g_i=g_\Phi$.
		\item With probability $(1-P_r)(1-B)$, select at random an agent
		$m \in \nu_i$ and set $g_i=g_m$.
		\item If $\Phi$ is autonomous, update the value $g_\Phi=\mode \{ g_i; \, i=1,2,\ldots,N\}$
	\end{enumerate}
	
	Step 2 specifies the rewiring process that modifies the network connectivity; new
	connections occur between agents with similar states. This rewiring decreases the number of links connecting nodes in different states, called active links. Links are rewired until a statistically stationary state, where the number of active links in the network drops to zero, is reached. Steps 3 and 4 comprise the node imitation dynamics of the voter model: step 3 expresses the agent–field interaction, while step 4 describes the agent–agent interaction. In the case of an autonomous global field $\Phi$, step 5 characterizes the time scale for the updating of state of the field $g_\Phi$. 
	We verified that the collective behavior of this system was statistically similar if the steps of the node dynamics were performed before the rewiring~process.

	\section{Results}
	
	In the absence of a global field ($B=0$), 
	the imitation dynamics of the nodes increases the number of connected agents with equal states, while the rewiring process favors the segregation and fragmentation of the network \cite{holme}. Therefore, the evolution of the system eventually leads to the formation of a set of separate components, or subgraphs, disconnected from each other, with all members of a subgraph sharing the same state. Such subgraphs are called {\it {domains}}. 
	
	To characterize the collective behavior of the coevolutionary system subject to a global field, we use, as an order parameter, the normalized  size of the largest domain in the system, averaged over several realizations of random initial conditions, denoted by $S_{\mbox{\scriptsize max}}$. 
	
		\begin{figure}[h]
		\includegraphics[scale=0.49]{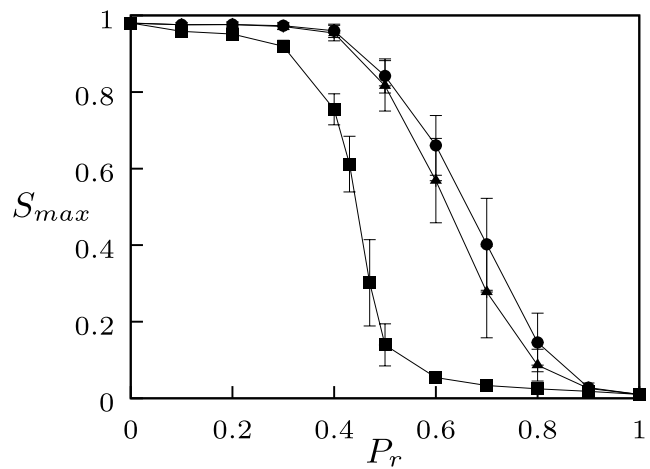}
		\caption{$S_{max}$ as a function of $P_r$ for the coevolutionary system subject to a global interaction field, for different values of the intensity $B$. The curves correspond to $B=0$ (squares); autonomous field with $B=0.003$ (triangles); external field with $B=0.003$ (circles). The parameters are $G=100$, $N=1000$, and  $\langle k \rangle=4$. The error bars indicate standard error obtained over $100$ realizations of random initial conditions for each value of $P_r$.}
		\label{f1} 
	\end{figure}

	Figure~\ref{f1} shows the quantity $S_{\mbox{\scriptsize max}}$ as a function of the rewiring probability $P_r$ for different values of the intensity of the global field $B$. When no global fields are present ($B=0$, squares), our model reduces to the coevolution model of Holme and Newman \cite{holme} where, as $P_r$ increased, $S_{\mbox{\scriptsize max}}$ exhibits a transition at the critical value  $P_r^*=0.458$, from a regime having a large domain whose size is comparable to the system size, characterized by values $S_{\mbox{\scriptsize max}} \rightarrow 1$, to a fragmented state consisting of small domains, for which $S_{\mbox{\scriptsize max}} \rightarrow 0$ \cite{holme}.

	Figure~\ref{f1} indicates that, when a global field either external or autonomous is applied to the system, the fragmentation transition persists, but the critical value of $P_r$ for which the transition takes place increases as $B$ increments. The one-large domain phase consists of agents sharing the state $g_F$ of the global field. For $B>0$, the critical value of $P_r$ in the presence of the external field is greater than the corresponding value for the autonomous field. As $B\rightarrow 1$, the fragmentation transition occurs at the value $P_r=1$ for either field. Thus, the presence of an external or an autonomous global field contributes to inhibit the fragmentation of the network. 
	
\begin{figure}[h]
		\includegraphics[scale=0.49]{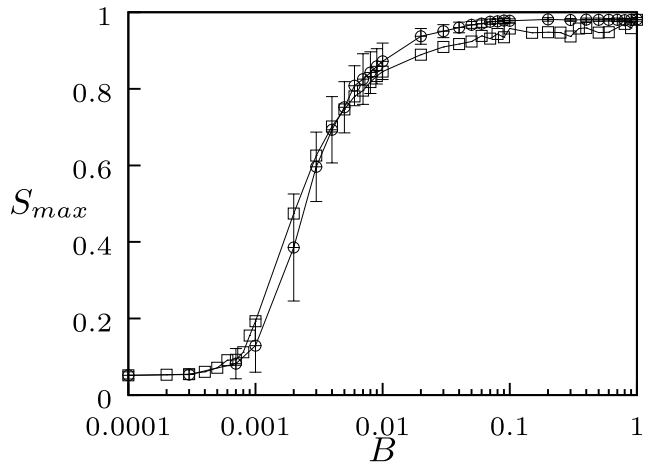}
		\caption{$S_{max}$ as a function of $B$ (log scale) with fixed $P_r=0.6>P_r^*$. The curves correspond to an autonomous field (open circles) and to an external field (open squares). The parameters are $G=100$, $N=1000$, and  $\langle  k \rangle=4$. The error bars indicate standard errors obtained over $100$ realizations of the initial conditions for each value of $B$.}
		\label{f2}
	\end{figure}

	In Figure~\ref{f2}, we show $S_{\mbox{\scriptsize max}}$ as a function of the intensity $B$, for both types of fields, with a fixed value of the rewiring probability $P_r=0.6>P_r^*$. For this value of $P_r$, the system reaches a fragmented state when $B = 0$. In Figure~\ref{f2}, we start from a fragmented state as the initial condition for each value of $B$. For small values of $B$, the system is fragmented in small domains, characterized by $S_{\mbox{\scriptsize max}} \rightarrow 0$. As the intensity $B$ increases above some critical value, both the external and the autonomous fields produce a recombination of the network: the small domains possessing multiple states become united into one large domain whose elements share the state $g_\Phi$ of the field. Thus, global interaction fields can have cohesive and homogenizing effects on a coevolutionary network.
	
\begin{figure}[h]
		\includegraphics[scale=0.49]{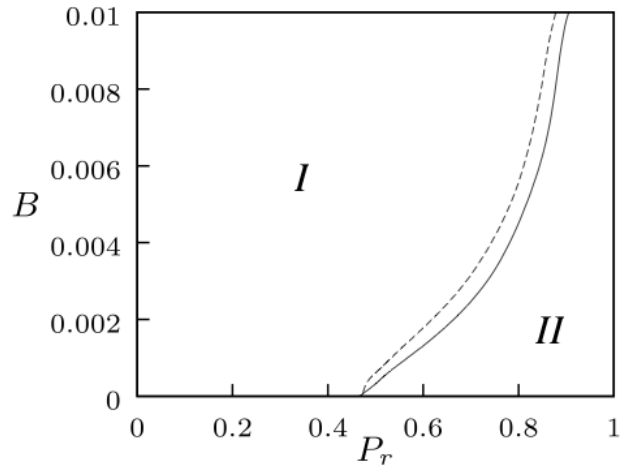}
		\caption{Critical boundaries for the fragmentation transition on the space of parameters $(B,P_r)$ for the adaptive system subject to an external global field (continuous line), or to an autonomous global field (dashed line). We calculate $S_{\mbox{\scriptsize max}}$ on the $(P_r,B)$ plane with resolutions of $10^{-2}$ for $P_r$ and $10^{-3}$ for $B$ and determine the critical values of $P_r$ and $B$ by using finite size scaling analysis \cite{holme}. In each case, the boundary separates two phases: ({\it I}) a phase with one-large domain sharing the state of the global field; and ({\it II}) a fragmented phase with many small domains.}
		\label{f3}
	\end{figure}

	The critical values of $B$ and $P_r$ for the fragmentation transitions in Figures~\ref{f1} and \ref{f2} were determined by using finite size scaling analysis, following the same approach proposed by Holme and Newman in \cite{holme}.
	Figure~\ref{f3} displays the collective behavior on the space of parameters $(B,P_r)$ for the coevolutionary system subject to an external global field and to an autonomous global field. In each case, a critical boundary separates two phases: ({\it I}) an ordered phase characterized by  $S_{\mbox{\scriptsize max}} \rightarrow 1$, where the network evolves to one large connected subgraph with all agents sharing the opinion of the field (above the curve); and ({\it II}) a fragmented phase for which $S_{\mbox{\scriptsize max}} \rightarrow 0$, where the system consists of many small subgraphs with different opinions. 
	An external global field appears a little more efficient than an autonomous field in preventing fragmentation in the coevolutionary system.
	
	\section{Discussion}
	We investigated the effects of the global interaction fields, external or autonomous,  on an adaptive network of social agents with an opinion formation dynamics. The phase diagram in Figure~\ref{f3} shows that the effects of both global fields on the collective behavior of the coevolutionary system were similar. The two phases arise from the competition between the homogenizing effect of the global field and the fragmentation of the network favored by the rewiring process. The similarity in the collective behaviors emerging in coevolutionary systems with external or autonomous global interaction fields signal that the nature of the field, either external or endogenous, is qualitatively irrelevant. At the local level, the field acts effectively as an additional influential neighbor for every agent with the same node dynamics in each case.
	
	In the context of dynamical systems, it has been shown that an analogy between an autonomous globally coupled system and a system subject to a global external drive can be established, because 
	all
	the elements in each of these systems are affected by the corresponding global field in the same way at a given time  \cite{Parra}. 
	Then, at the local level in either system, each element can be described as a drive--response dynamical system that can synchronize, and which eventually manifests as a collective state of synchronization. In social dynamics, a state of consensus can be interpreted as synchronization. In particular, for the coevolutionary opinion formation model considered, either external mass media or endogenous mass media trends induce consensus about their respective state. Our results suggest that the analogy between dynamical systems possessing external or autonomous global interaction fields can be extended to coevolutionary systems subject to global fields.
	
	The opinion dynamics based on the simple imitation employed in this model led to the imposition of the states of the global mass media fields on the system. The imitation rule of voter dynamics has been applied to model elections, language competition, and clustering processes.
	We have found that a global field, either external or autonomous, can induce the recombination of a network broken in small domains into one large domain. In a social context, external mass media as well as the feedback of endogenous cultural trends can play a major role in preventing fragmentation and favoring cohesion in a society that possesses coevolution dynamics. Global mass media may contribute to control voters polarization and segregation in adaptive social networks. On the other hand, even under the influence of mass media, the existence of adaptive rewiring may counter the expected consensus and cohesion, as phase II on the phase diagram in Figure~\ref{f3} shows.
	
	Global mass media acting on systems possessing non-interacting states in their dynamics, such as in Axelrod's model for cultural dissemination \cite{Axelrod} or Deffuant’s bounded confidence model \cite{Deff}, can produce nontrivial effects other than imposing consensus, such as inducing disorder \cite{Media1},
	alternative ordering  \cite{NJP}, the emergence of chimera states~\cite{Cross}, and promoting minority growth and polarization \cite{Media2}. Future interesting extensions to be investigated include the influence of different node dynamics, such as those with non-interacting states, on the collective behavior of coevolutionary systems subject to  global fields, general coevolution models \cite{Herrera}, and the characterization of the topological properties of adaptive networks driven by global fields.

\vspace{9mm}

	\section*{Acknowledgment}
	This work was supported by Corporaci\'on Ecuatoriana para el Desarrollo de la
		Investigaci\'on y Academia (CEDIA) through grant
		CEPRA XVI-2022-09 ``Desarrollo y Aplicaciones de Recursos Computacionales en Sociof\'isica".


\begin{thebibliography}{999}
			\bibitem{Kaneko} Kaneko, K. Chaotic but regular posi-nega switch among coded attractors by cluster-size variation. \emph{Phys. Rev. Lett.} \textbf{1989}, \emph{63}, 219.
			
			\bibitem{Parra}  Cosenza, M.G.; Parravano, A. Dynamics of coupling functions in globally coupled maps: Size, periodicity, and stability of clusters.
			\emph{Phys. Rev. E} \textbf{2001}, \emph{64}, 036224.
			
			\bibitem{Piko} Pikovsky, A.; Rosenblum, M. Dynamics of globally coupled oscillators: Progress and perspectives. \emph{Chaos} \textbf{2015}, \emph{25}, 097616.
			
			\bibitem{Kuramoto} Kuramoto, Y. \textit{Chemical Oscillations, Waves and Turbulence}; Springer: Berlin/Heidelberg, Germany,1984.
			
			\bibitem{Naka} Nakagawa, N.; Kuramoto, Y. From collective oscillations to collective chaos in a globally coupled oscillator system. \emph{Physica D} \textbf{1994}, \emph{75}, 74.
			
			
			\bibitem{Wie} Wiesenfeld, K.; Hadley, P. Attractor crowding in oscillator arrays. \emph{Phys. Rev. Lett.} \textbf{1989}, \emph{62}, 1335.
			
			\bibitem{Gruner} Gr\"uner, G. The dynamics of charge-density waves. \emph{Rev. Mod. Phys.} \textbf{1988}, \emph{60}, 1129.
			
			\bibitem{Wie2} Wiesenfeld, K.; Bracikowski, C.; James, G.; Roy, R.
			Observation of antiphase states in a multimode laser. 
			\emph{Phys. Rev. Lett.} \textbf{1990}, \emph{65}, 1749.
			
			\bibitem{Kan1} Kaneko, K.; Tsuda, I. \textit{Complex Systems: Chaos and Beyond}; Springer: Berlin/Heidelberg, Germany,
			2001.
			
			\bibitem{Newman} Newman, M.; Barab\'asi, A.; Watts, D.J. \textit{The Structure and
				Dynamics of Networks}; Princeton University Press, Princeton,
			NJ, USA, 2006.
			
			\bibitem{Yako} Yakovenko, V.M. \textit{Encyclopedia of Complexity and System
				Science}; Meyers, R.A., Ed.; Springer: New York, NY, USA, 2009.	
			
			\bibitem{Media1} Gonz\'alez-Avella, J.C.; Cosenza, M.G.; Tucci, K. Nonequilibrium transition induced by mass media in a model for social influence. \emph{Phy. Rev. E} \textbf{2005}, \emph{72}, 065102.
			
			\bibitem{NJP}  Gonz\'alez-Avella, J.C.; Cosenza, M.G.; Eguiluz, V.M.; Miguel, M.S. Spontaneous ordering against an external field in non-equilibrium systems. \emph{New J. Phys.} \textbf{2010}, \emph{12}, 013010.
			
			\bibitem{Cross} Gonz\'alez-Avella, J.C.; Cosenza, M.G.; Miguel, M.S.
			Localized coherence in two interacting populations of social agents. \emph{Physica A} \textbf{2014}, \emph{399}, 24.
			
			\bibitem{Media2} Cosenza, M.G.; Gavidia, M.E.; Gonz\'alez-Avella, J.C. 
			Against mass media trends: Minority growth
			in cultural globalization. \emph{PloS One} \textbf{2020} \emph{15}, e0230923.
			
			\bibitem{Kan2} Kaneko, K. Clustering, coding, switching, hierarchical ordering, and control in a network of chaotic elements. \emph{Phys. D} \textbf{1990}, \emph{41}, 137.
			
			\bibitem{Manrubia} Manrubia, S.C.; Mikhailov, A.S.; Zanette, D.H. \textit{Emergence of Dynamical Order: Synchronization Phenomena in Complex Systems}; World Scientific: Singapore, 2004.
			
			\bibitem{Sethia} Sethia, G.C.; Sen, A. Chimera states: The existence criteria revisited. \emph{Phys. Rev. Lett.} \textbf{2014}, \emph{112}, 144101.
			
			\bibitem{Yedel} Yeldesbay, A.; Pikovsky, A.; Rosenblum, M. Chimeralike states in an ensemble of globally coupled oscillators. \emph{Phys. Rev. Lett.} \textbf{2014}, \emph{114}, 144103.
			
			\bibitem{Ojalvo}Garcia-Ojalvo, J.; Elowitz, M.B.; Strogatz, S.H. Modeling a synthetic multicellular clock: Repressilators coupled by quorum sensing. \emph{Proc. Natl.
				Acad. Sci. USA} \textbf{2004}, \emph{101}, 10955.
			
			\bibitem{Wang} Wang, W.; Kiss, I.Z.; Hudson, J.L. Experiments on arrays of globally coupled chaotic electrochemical oscillators: Synchronization and clustering. \emph{Chaos} \textbf{2000}, \emph{10}, 248.
			
			\bibitem{Monte} Monte, S.D.; d’Ovidio, F.; Dano, S.; Sorensen, P.G.
			Dynamical quorum sensing: Population density encoded in cellular dynamics.
			\emph{Proc. Natl. Acad. Sci. USA} \textbf{2007}, \emph{104}, 18377.
			
			\bibitem{Taylor} Taylor, A.F.; Tinsley, M.R.; Wang, F.; Huang, Z.; Showalter, K. Dynamical quorum sensing and synchronization in large populations of chemical oscillators.
			\emph{Science} \textbf{2009}, \emph{323}, 614.
			
			\bibitem{Tinsley} Tinsley, M.R.; Nkomo, S.; Showalter, S. Chimera and phase-cluster states in populations of coupled chemical oscillators. \emph{Nat. Phys.} \textbf{2012}, \emph{8}, 662.
			
			\bibitem{Scholl} Hagerstrom, A.M.; Murphy, T.E.; Roy, R.; H\"ove, P.; Omelchenko, I.; Sch\"oll, E. Experimental observation of chimeras in coupled-map lattices. \emph{Nat. Phys.} \textbf{2012}, \emph{8}, 658.
			
			\bibitem{Kan3} Ouchi, N.B.; Kaneko, K. Coupled maps with local and global interactions. \emph{Chaos} \textbf{2000}, \emph{10}, 359.
			
			\bibitem{Eguiluz1} Zimmermann, M.G.; Eguiluz, V.M.; Miguel, M.S.; Spadaro A. Cooperation in an adaptive network. \emph{Adv. Complex Syst.} \textbf{2000}, \emph{3},  283.
			
			\bibitem{Rohl} Bornholdt, S.; Rohlf, T. Topological evolution of dynamical networks: Global criticality from local dynamics. \emph{Phys. Rev. Lett.} \textbf{2000}, \emph{84}, 6114.
			
			\bibitem{Eguiluz2} Zimmermann, M.G.; Eguiluz, V.M.; ; Miguel, M.S. Coevolution of dynamical states and interactions in dynamic networks. \emph{Phys. Rev. E} \textbf{2004}, \emph{69},  065102.
			
			\bibitem{Blasius} Gross, T.; Blasius, B. Adaptive coevolutionary networks: a review. \emph{J. R. Soc. Interface} \textbf{2008}, \emph{5}, 
			259.
			
			\bibitem{Sayama} Gross, T.; Sayama, H. \emph{Adaptive Networks: Theory, Models, and Applications}; Springer: Berlin/Heidelberg, Germany,
			2009.
			
			\bibitem{Ito} Ito, J.; Kaneko, K. \emph{Neural Netw.} Self-organized hierarchical structure in a plastic network of chaotic units. \textbf{2000}, \emph{13}, 275.
			
			\bibitem{Gross} Meisel, C.; Gross, T. Adaptive self-organization in a realistic neural network model. \emph{Phys. Rev. E} \textbf{2009}, \emph{80}, 
			061917.
			
			\bibitem{Ito2} Ito, J.; Kaneko, K. Spontaneous structure formation in a network of chaotic units with variable connection strengths. \emph{Phys. Rev. Lett.} \textbf{2002}, \emph{88}.
			028701.
			
			\bibitem{Gong}  {Gong, P.; Van Leeuwen}, C. Evolution to a small-world network with chaotic units. \emph{Europhys. Lett.} \textbf{2004}, \emph{67},
			328.
			
			\bibitem{Shibata} Shibata, T.; Kaneko, K. Coupled map gas: structure formation and dynamics of interacting motile elements with internal dynamics. \emph{Phys. D} \textbf{2003}, \emph{181}, 197.
			
			\bibitem{Arenas} Arenas, A.; D\'iaz-Guilera, A.; Kurths, J.; Moreno, Y.; Zhou, C. Synchronization in complex networks. \emph{Phys. Rep.} \textbf{2008}, \emph{469}, 93.
			
			\bibitem{Mandra} Mandr\'a, S.; Fortunato, S.; Castellano, C. Coevolution of Glauber-like Ising dynamics and topology. \emph{Phys.
				Rev. E} \textbf{2009}, \emph{80}, 056105.
			
			\bibitem{Gross2} Gross, T.; D’Lima, C.D.; Blasius, B. Epidemic dynamics on an adaptive network. \emph{Phys.
				Rev. Lett.} \textbf{2009}, \emph{96}, 208701.
			
			\bibitem{Risau} Risau-Gusman, S.; Zanette, D.H. Contact switching as a control strategy for epidemic outbreaks. \emph{J. Theor. Biol.}
			\textbf{2009}, \emph{257}, 52.
			
			
			\bibitem{Shaw} Schwartz, I.B.; Shaw, L.B. Rewiring for adaptation. \emph{Physics} \textbf{2010}, \emph{3}, 17.
			
			\bibitem{Gao} Gao, J.; Li, Z.; Wu, T.; Wang, L. The coevolutionary ultimatum game. \emph{EPL (Europhys. Lett.)} \textbf{2011}, \emph{93}, 48003.
			
			\bibitem{holme} Holme, P.; Newman, M.E.J. Nonequilibrium phase transition in the coevolution of networks and opinions. \emph{Phys. Rev. E} \textbf{2006}, \emph{74},
			056108.
			
			\bibitem{Min1} Min, B.; Miguel, M.S. Fragmentation transitions in a coevolving nonlinear voter model. \emph{Sci. Rep.} \textbf{2017}, \emph{7}, 12864.
			
			\bibitem{20}  Centola, D.; Gonz\'alez-Avella, J.C.; Eguiluz, V.M.; Miguel, A.M.S. Homophily, cultural drift, and the co-evolution of cultural groups. \emph{J. Conf. Res.} \textbf{2007}, \emph{51}, 905.
			
			\bibitem{22} Kozma, B.; Barrat, A. Consensus formation on adaptive networks. \emph{Phys. Rev. E} \textbf{2008}, \emph{77}, 016102.
			
			\bibitem{23} Kimura, D.; Hayakawa, Y. Coevolutionary networks with homophily and heterophily. \emph{Phys. Rev. E} \textbf{2008}, \emph{78}, 016103.
			
			\bibitem{24} Medo, M.; Zhang, Y.C.; Zhou, T. Adaptive model for recommendation of news. \emph{EPL (Europhys. Lett.)}  \textbf{2009}, \emph{88}, 38005.
			
			\bibitem{Herrera} Herrera, J.L.; Cosenza, M.G.; Tucci, K.; Gonz\'alez-Avella, J.C. General coevolution of topology and dynamics in networks.
			\emph{EPL (Europhys. Lett.)} \textbf{2011}, {95}, 58006.
			
			\bibitem{Gargi} Gargiulo, F.; Huet, S. Opinion dynamics in a group-based society. \emph{EPL (Europhys. Lett.)} \textbf{2010}, \emph{91}, 58004.
			
			\bibitem{Network} Estrada, E.; Knight, P.A. \textit{A First Course in Network Theory}; Oxford University Press: Oxford, UK, 2015.
			
			\bibitem{Axelrod} Axelrod, R. The dissemination of culture: A model with local convergence and global polarization. \emph{J. Conflict Res.}  \textbf{1997}, \emph{41}, 203.
			
			\bibitem{Deff} Deffuant, G.; Neau, D.; Amblard, F.; Weisbuch, G. Mixing beliefs among interacting agents. \emph{Adv. Complex Sys.} \textbf{2000}, \emph{3}, 87.
			
		\end{thebibliography}
\end{document}